\documentclass[12pt]{iopart}
\usepackage{iopams}
\usepackage{graphicx,cite,epsf}
\usepackage{psfrag}
\usepackage{ulem}
\RequirePackage{color}
\definecolor{MyDarkGreen}{rgb}{0.02,0.60,0.06}

\begin{document}

\title[Size and shape properties of ring polymers on percolation clusters]{Ring polymers on percolation clusters}

\author{ K. Haydukivska$^{1,2}$ and V. Blavatska$^{1,2}$ }
\address{$^1$ Institute for Condensed Matter Physics of the National Academy of Sciences of Ukraine,
79011 Lviv, Ukraine}
\address{$^2$ $\mathbb{L}^4$ Collaboration $\&$ Doctoral College for the Statistical Physics of Complex Systems, Leipzig-Lorraine-Lviv-Coventry, Europe}

\begin{abstract}
In the present work,  the cyclic polymer chains (rings) in structurally disordered environment (e.g. in the cross-linked polymer gel) are studied exploiting the model of  closed self-avoiding walks (SAWs) trajectories on $d=3$-dimensional percolation clusters.
  Numerical simulations with an application of pivot algorithm are performed. The estimates for the universal size and shape characteristics such as size ratios, averaged asphericity and prolateness of typical polymer conformation are obtained. Our results quantitatively describe an elongation and increase of anisotropy of ring polymers in disordered environment comparing with  the pure solvent.
\end{abstract}
\date{\today}

\submitto{Journal of Physics: Condensed Matter}
\section{Introduction}

Closed polymer chains (rings) gain significant interest of researchers with
 the discovery of certain DNA in
circular form \cite{Fiers62,Zhou03,Wasserman86}. Under
appropriate synthesis conditions, the  covalent  linking of the the ends of
flexible linear chains (ring formation)
 is observed during polymerization and polycondensation \cite{Brown65,Geiser80,Roovers83}.
The statistical properties of circular polymers in dilute solutions have been intensively studied both analytically \cite{Bishop85,Bishop86a,Bishop86b,Bishop88,Diel89,Jagodzinski92,Obukhov94,Muller99,Deutsch99,Miyuki01,Calabrese01,Alim07,Bohr10,Sakaue11,
Jung12,Ross11}
and experimentally \cite{Roovers83,Higgins79}.

In statistical description of long flexible polymers  in solvents  one can distinguish a number of  conformational properties, which possess universality, i.e. are  independent of details of chemical structure, and thus allow to combine the wide range of macromolecules  into the so called universality classes.
As typical examples, one can consider  the averaged radius of gyration $\langle R_{g\,{\rm chain}}^2\rangle$ and the end-to-end distance $\langle R_{e\,{\rm chain}}^2\rangle$ of  linear polymer chains,  obeying the scaling law with number of monomers $N$ \cite{deGennes,desCloiseaux}:
\begin{equation}\label{scalingR}
  \langle R_{g\,{\rm chain}}^2 \rangle \sim \langle R_{e\,{\rm chain}}^2 \rangle
\sim N^{2\nu}, \label{scalR}
\end{equation}
where $\langle (\ldots) \rangle$ means averaging over an ensemble of possible conformations of macromolecule and $\nu$ is universal critical exponent,  depending on the space dimension $d$.
 In particular, a nice approximation is given by Flory formula \cite{deGennes}: $\nu(d)=3/(d+2)$, which restores the fact
 that at space dimension above the upper
critical one $d_{\rm up}{=}4$ one has an ideal Gaussian polymer with $\nu (d\geq4){=}1/2$  (apart from logarithmic  corrections to scaling). The value of critical exponent $\nu$ in \ref{scalR} is not influenced by changing the topology of polymer structure, and thus the radius of gyration  $\langle R_{g\,{\rm ring}}^2\rangle$ of closed polymer rings obeys exactly the same scaling law \cite{Duplantier94,Prentis82,Privman}. Thus, the useful parameter to compare the  size measures of linear and ring polymers of the same molecular weight $N$ have been introduced \cite{Zimm49}:
 \begin{equation} \label{g}
 g\equiv \frac{\langle R_{g\,{\rm ring}}^2\rangle}{\langle R_{g\,{\rm chain}}^2\rangle},
 \end{equation}
 which is in turn  universal and $N$-independent quantity. In the case of  idealized  Gaussian case one has  $g=1/2$  \cite{Zimm49},
whereas presence of excluded volume effect leads to an increase of this value (see Table 1).
Note that for the closed circular polymers, the spanning radius  $ R_{1/2\,{\rm ring}} $ is of interest instead of the usual end-to-end distance, and  a ratio
\begin{equation} \label{p}
 p\equiv \frac{\langle R_{1/2\,{\rm ring}}^2\rangle}{\langle R_{g\,{\rm ring}}^2\rangle}
 \end{equation}
is introduced. For a Gaussian polymer one has $p=3$, and this ratio again increases when excluded volume is taken into account \cite{Prentis82}.

\begin{table}[t!]
\begin{tabular}{| c | c | c | c | c |}
   \hline
$$    &  $g$                                  & $p$                                & ${\overline{ \langle{A_3} }\rangle}$                               & ${\overline{ \langle{P} \rangle}}$ \\ \hline
pure   &  $0.536(7)$\cite{Jagodzinski92}  & $3.217(20)$\cite{Jagodzinski92}& $0.262(1)$\cite{Jagodzinski92} & $0.205(2)$\cite{Jagodzinski92}\\
      &      $0.539(25)$  \cite{Bishop88}          &                 & $0.255(10)$\cite{Bishop86a}    &        $0.151 (14)$ \cite{Bishop86a}         \\
&  $0.516$\cite{Douglas84} & & & \\
&  $0.53(3)$\cite{Higgins79} & & & \\
   our study   &  $0.537(1)$                      & $3.288(3)$                    & $0.258(1)$                     & $ 0.198(1)$ \\ \hline
$pc$,

 our study&  $0.548(2)$                      & $3.463(6)$                    & $0.264(1)$                     & $0.201(1)$\\ \hline
\end{tabular}
\caption{Size ratios and shape parameters of ring polymers in $d=3$  in pure environment and on the percolation cluster.}\label{t1}
\end{table}

To characterize the shape properties of typical polymer chain conformation, the  rotationally invariant universal quantities, such as the averaged asphericity
$\langle A_d \rangle$ and prolateness $\langle P \rangle$ have been introduced \cite{Aronovitz86,Rudnick}. $\langle A_d \rangle$ takes on a maximum value of one for a completely stretched, rod-like configuration,
and equals zero for spherical form, thus obeying the inequality:  $0\leq \langle A_d \rangle \leq 1$. Prolateness $\langle P \rangle$ is positive for prolate ellipsoid-like configurations, and is negative for oblate shapes, being bounded to
the interval $-1/4 \leq  \langle  P \rangle \leq 2$.  For a Gaussian ring, an exact value for asphericity parameter is obtained $ \langle A_d \rangle =(d+2)/(5d+2)$ \cite{Gaspari87} . The values of $\langle A_d \rangle$ and $\langle P \rangle$
for ring polymers with excluded volume effect are given in Table 1.

 The model of  self-avoiding walk (SAW) on a regular lattice  was proven to be very successful in
in capturing  the universal configurational  properties of  flexible polymers in good solvent \cite{desCloiseaux}. The ring polymer can thus be straightforwardly modeled as a closed SAW trajectory, where the first and the last steps coincide. This model have been exploited in numerous numerical studies of circular polymers \cite{Bishop85,Bishop86a,Bishop86b,Bishop88} and  allows to  obtain the reliable quantitative estimates  for their size and shape properties.

\begin{figure}[b!]
\begin{center}
\includegraphics[width=130mm]{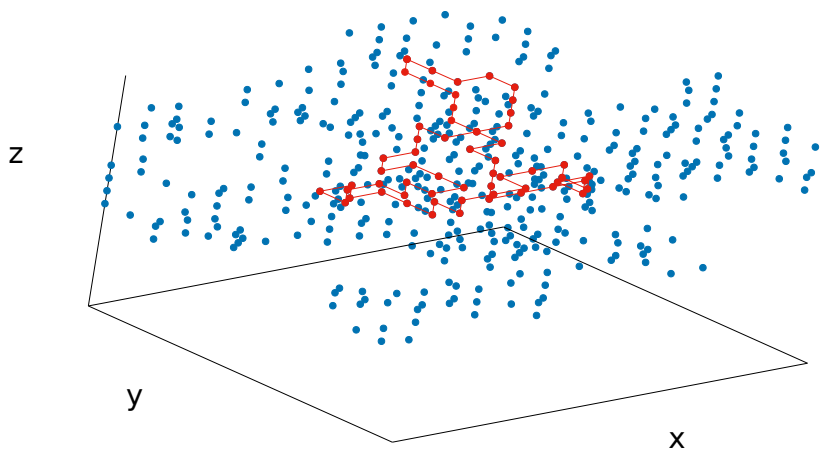}
\caption{(Color online): The part of  percolation cluster constructed on cubic lattice and closed SAW trajectory, allowed to take its steps only on sites belonging to percolation cluster.\label{pcring} }
\end{center}
\end{figure}

There are numerous physical situations, in which the ring polymers are placed in an
structurally disordered  environment, e.g. the cross-linked polymer gel in processes of  gel electrophoresis \cite{Dorfman},
colloidal solutions \cite{Pusey86},  intra- and extracellular environments \cite{Kumarrev} which are highly crowded due to the presence of a large amount of various biochemical species \cite{Minton01}.
An important question  is how the universal conformational properties of macromolecules are modified in presence of structural
obstacles (impurities) in the system.
Within the frames of lattice model of polymers,  a disordered medium can be modeled as a randomly diluted lattice,
where only a given concentration $p$ of randomly chosen lattice sites allowed for SAWs,  whereas the remaining $1-p$ sites containing point-like obstacles.
 Mostly interesting is the case, when concentration
$p$ equals the threshold value $p_{c}$ (in $d=3$, in the case of site percolation it reads $p_{c}=0.31160$ \cite{Grassberger92})
and an incipient percolation cluster can be found in the system \cite{Stauffer}.
The complicated fractal structure of percolation cluster
captures the effect of density fluctuations of structural defects, which often lead a considerable spatial inhomogeneity and create pore spaces of fractal structure \cite{Dullen79}.
In particular, it can be used in modeling the gel structure, which is highly irregular and contain multiple dangling ends \cite{Whytock}, which is a distinctive feature of percolation cluster.

  Numerous analytical and numerical studies \cite{Woo91,Grassberger93,Rintoul94,Ordemann02,Janssen07,Blavatska10,Blavatska10a} quantitatively confirmed the extension of effective size and an increase of anisotropy of typical linear polymer chain conformations on percolation cluster. It is important to note,
that complex fractal  structure  of underlying percolation cluster causes the change of universality class of  SAWs residing on it. In particular,
the scaling law (\ref{scalR}) holds in this case with larger value of critical exponent $\nu_{p_c}(d=3)=
0.667(3)$ \cite{Blavatska10}, and
asphericity and prolateness values considerably increase comparing with the case of pure lattice  \cite{Blavatska10a}.
However, though the statistical properties of circular polymers in diluted environments have been considered in numerical simulations \cite{Gersappe94, Kuriata15,Michieletto15}, the
thorough study of  universal size and shape characteristics of this polymer topology on percolation cluster have not been performed so far and is the aim of the present study.

The layout of the paper is as follows. In the next Section, we short describe the methods of constructing of underlying 
percolation cluster on cubic lattice and pivot algorithm, used for simulating SAW. The Section \ref{results} contain our main simulation results on size and shape characteristics of closed SAW on percolation cluster. Conclusions are given in Section \ref{conc}.

\section{The method}

\subsection{Construction of the percolation cluster}

We consider site percolation on a regular 3-dimensional lattice of edge length $L=100$. Each site of the lattice is considered to be
empty with probability $p_c=0.31160$, and containing a point-like obstacle otherwise.
To extract the percolation cluster, we apply an algorithm developed
by Hoshen and Kopelman \cite{Hoshen76}. As a first step, a label is prescribed to each  empty site in increasing order.
 At the second step,  for each of the labeled site (say, $i$), we check whether its nearest neighbors are
also empty. If the  label of the neighbor is larger than that of  site $i$, we change the label of the  neighbor to that of site $i$.
Otherwise, we change the label of site $i$ to that of the neighbor.
Such an algorithm is applied until no more changes of site labels are needed. As a result, we  obtain the  groups of clusters of empty sites of different sizes, where all the sites in a given group have the same label.   Finally, we check for the existence of  a cluster,
that wraps around the lattice. If it exists, we have found the percolation cluster. If not, this disordered lattice is rejected
and a new one is constructed.
 A total number  $C=1400$ clusters have been constructed.

\subsection{ Pivot Algorithm}

\begin{figure}[b!]
\begin{center}
\includegraphics[width=40mm]{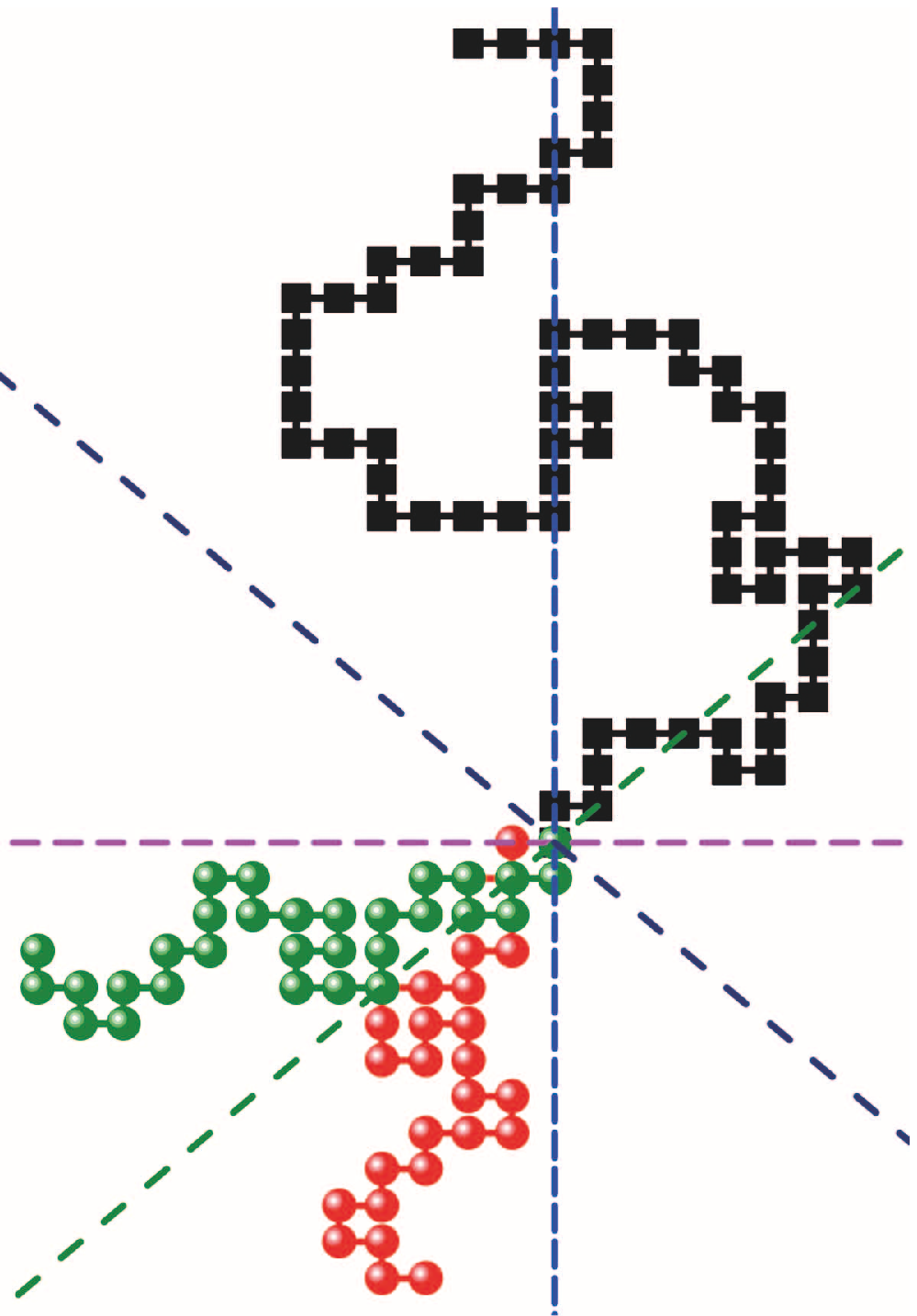}\qquad\includegraphics[width=20mm]{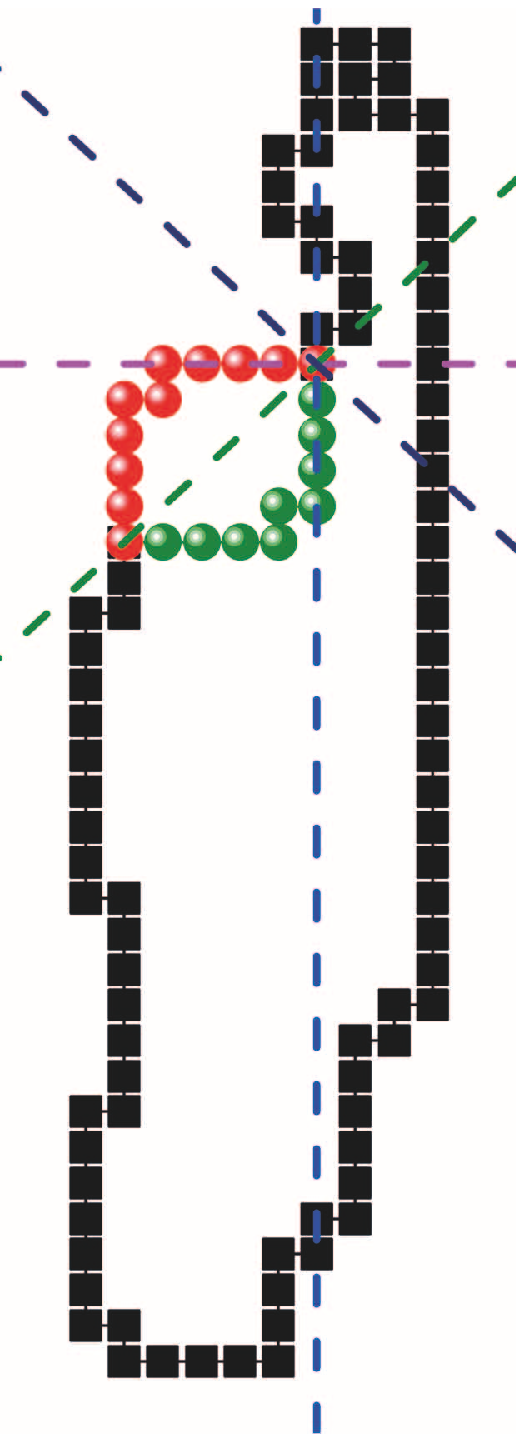}
\caption{(Color online): Schematic presentation of one of the pivot operations for an open  (left) and closed (right) SAW trajectory. Black symbols represent part of the walk that remains intact, red symbols represent parts that get transformed, while green ones depict the results of the transformation.
Dash lines show axis of symmetry.  \label{op} }
\end{center}
\end{figure}

To study the SAWs having their steps only on the sites belonging to percolation cluster (see Fig. \ref{pcring}), we apply the  pivot algorithm, proposed in Ref.\cite{Lal}
and further developed in Refs. \cite{clisby10,madras88} among others. At a first step,  an initial configuration of  SAW trajectory of a given length $N$ is build.
 The SAW is considered to be closed when a walker of $N$ steps returns to the starting point.
Then, the sequence of elementary pivot moves, i.e. a lattice symmetry operations (rotation or reflection) are
applied to parts of a trajectory.
When dealing with open chains,  we choose a random point on this trajectory so that it is split into two parts. Then a randomly chosen
symmetry operation is applied to one of the parts. The result of this operation is accepted if the resulting walk is also a self-avoiding and all its steps are
on the sites belonging to  percolation cluster. If not, the initial configuration is considered one more time.
 Then a new point is chosen and the procedure is repeated $P$-times.
When dealing with closed trajectories, one more step has to be added in algorithm to ensure that the resulting configuration is also closed.
For each of the symmetry operations there  always is a plane or an axis of symmetry associated with it and a second point on the closed SAW trajectory
that also belongs to that plane or axis (see fig. \ref{op}) can be chosen, so that pivot step can be performed between two points.

We consider both open and closed SAW trajectories with the lengths up to $N=40$ and perform up to $P=20000$ and $P=10000$ pivot steps correspondingly.
The lower amount of pivot steps for closed SAWs is considered because there is a smaller amount of configurations available comparing with the open case.
This procedure is performed on all percolation clusters constructed,  taking into account up to $30$  different starting configurations on each cluster.

The estimates for any observable $O$ is obtained by performing  the double
 averaging: the first one is performed over all SAW configurations on a single
percolation cluster
\begin{equation}
\langle O \rangle =  \frac{1}{P}\sum_{i=1}^P O_i,
\end{equation}
the second average is carried out over different
realizations of disorder, i.e. over all percolation clusters constructed:
\begin{equation}
\overline{\langle O \rangle} = \frac{1}{C}\sum_{j=1}^C \langle O \rangle_j.
\end{equation}

\section{Results}\label{results}

\begin{figure}[b!]
\begin{center}
\includegraphics[width=73mm]{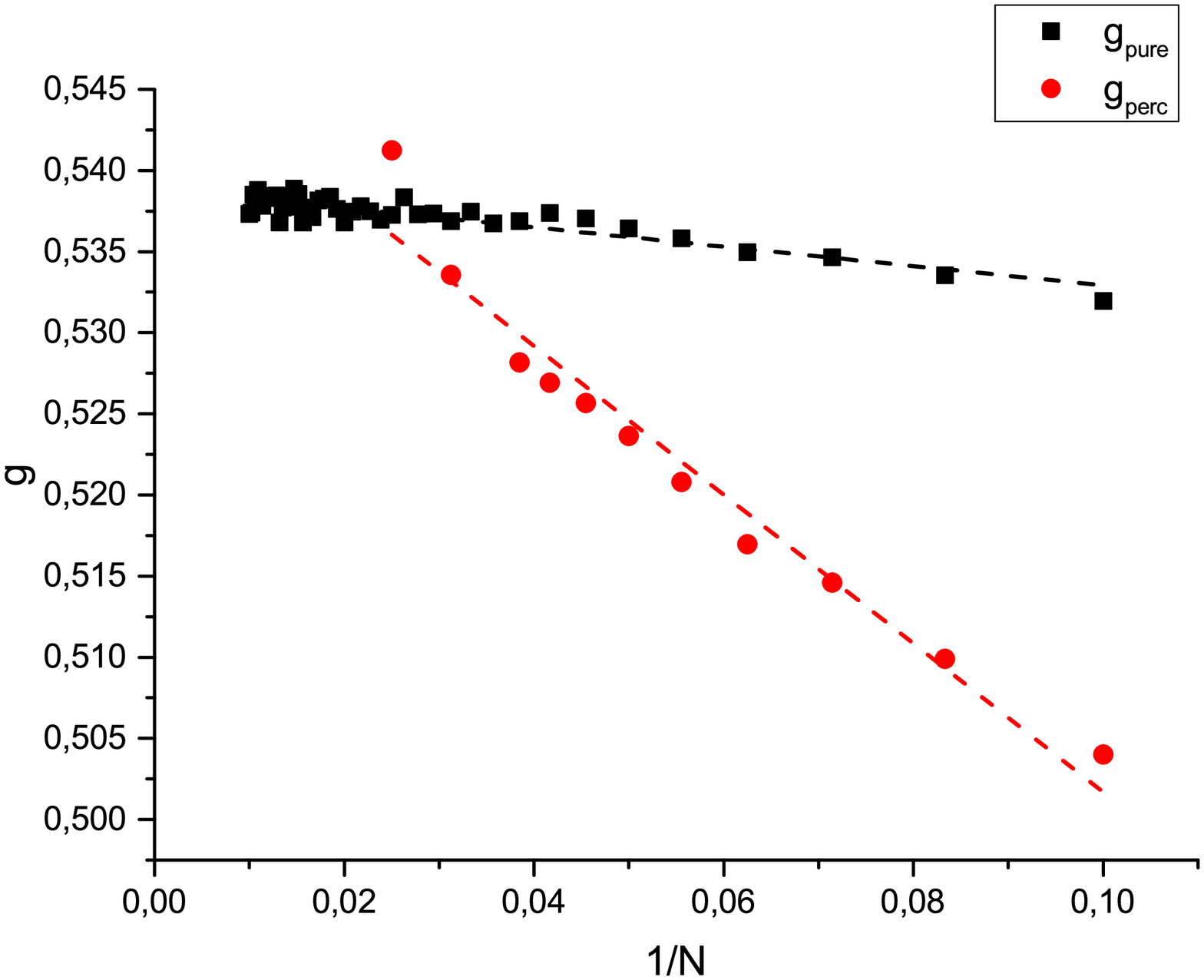}\includegraphics[width=73mm]{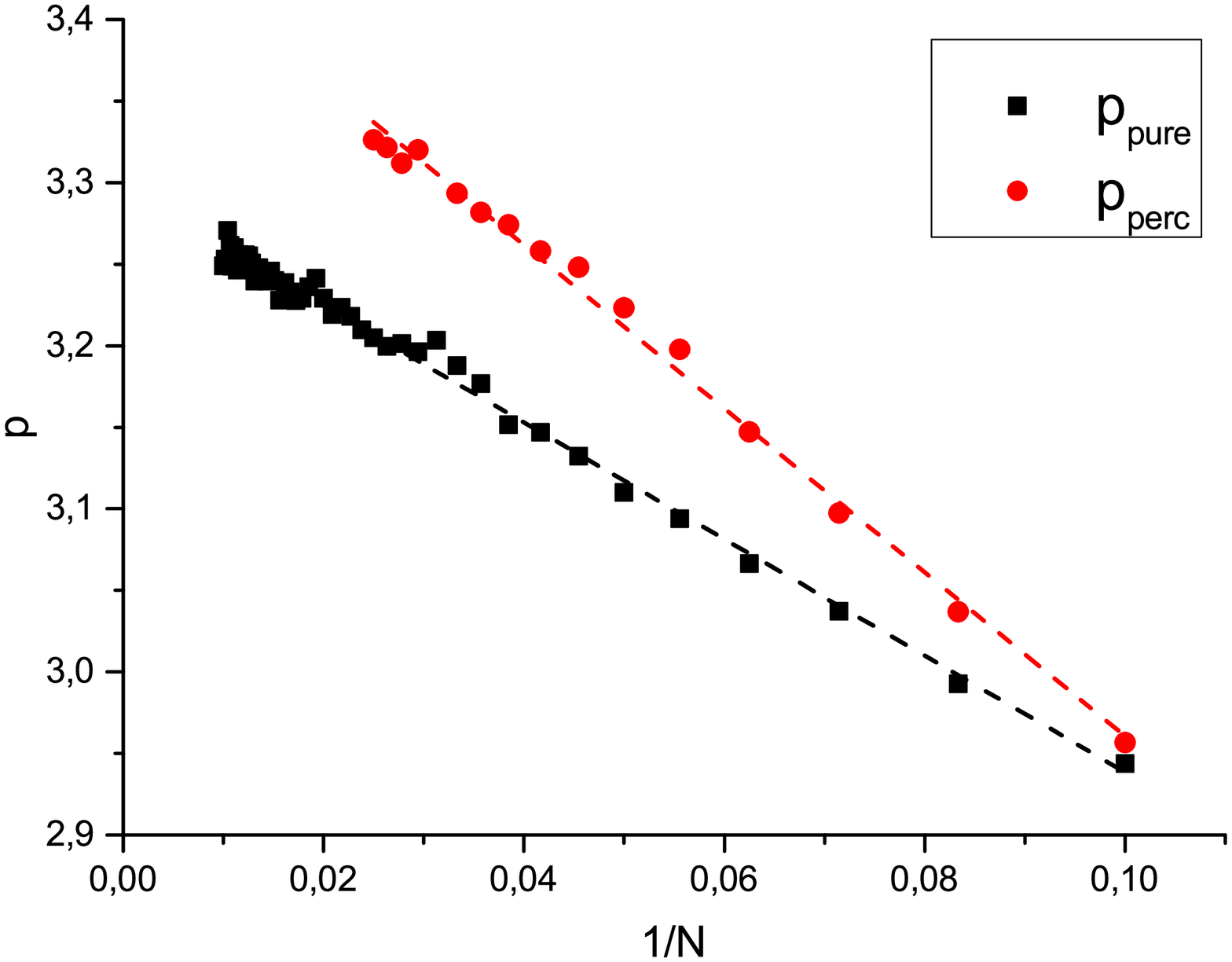}
\caption{ \label{fig:1} Size ratio (\ref{g}) on the left and  size ratio (\ref{p}) on the right as  functions of $1/N$.
Black squares: pure lattice, red circles:  percolation cluster.}
\end{center}
\end{figure}

To find the estimates for size and shape parameters,  we start with evaluation of the components of gyration tensor $\bf Q$, which  in $d=3$ are given by \cite{Aronovitz86,Rudnick}:
\begin{equation}
Q_{ij}=\frac{1}{N}\sum_{n=1}^N(x_n^i-{x^i_{CM}})(x_n^j-{x^j_{CM}}),\,\,\,\,\,\,i,j=1,\ldots,3,
\label{mom}
\end{equation}
with $\{x_n^{1},\ldots,x_n^3\}$ being the set of coordinates of  position vector  $\vec{R}_n$ of the $n$th step of SAW trajectory ($n=1,\ldots,N$)
and  ${x^i_{CM}}=\sum_{n=1}^Nx_n^i/N$  the coordinates of the center-of-mass position vector ${\vec{R}_{CM}}$.
 The squared radius of gyration is given by:
\begin{equation}
R_g^2 =\frac{1}{N} \sum_{n=1}^N (\vec{R}_n-{\vec{R}_{CM}})^2  =  \sum_{i=1}^d Q_{ii} = {\rm Tr}\, \bf{Q}, \label{rg}
\end{equation}
whereas asphericity ${A_d}$ and prolateness $P$ are defined in terms of components of gyration tensor according to \cite{Aronovitz86}:
\begin{eqnarray}
{A_3} =\frac{1}{6} \sum_{i=1}^3\frac{(\lambda_{i}-{\overline{\lambda}})^2}{\overline{\lambda}^2}=
\frac{3}{2}\frac{\rm {Tr}\,{\bf{\hat{Q}}}^2}{(\rm{Tr}\,{\bf{Q}})^2}, \label{add}\\
P =\frac{\prod_{i=1}^3(\lambda_{i}-{\overline{\lambda}})}{{\overline{\lambda}}^3}=27\frac{{\rm det}\, \bf{{\hat{Q}}}}{(\rm{Tr}\,{\bf{Q}})^3}, \label{sdd}
\end{eqnarray}
with $\lambda_i$ being the spread of eigenvalues of gyration tensor, ${\overline{\lambda}}\equiv {\rm Tr}\, {\bf{Q}}/d$
 the average eigenvalue and  ${\bf{{\hat{Q}}}}\equiv{\bf{Q}}-\overline{\lambda}\,{\bf{I}}$ (here $\bf{I}$ is  the unity matrix).

\begin{figure}[t!]
\begin{center}
\includegraphics[width=73mm]{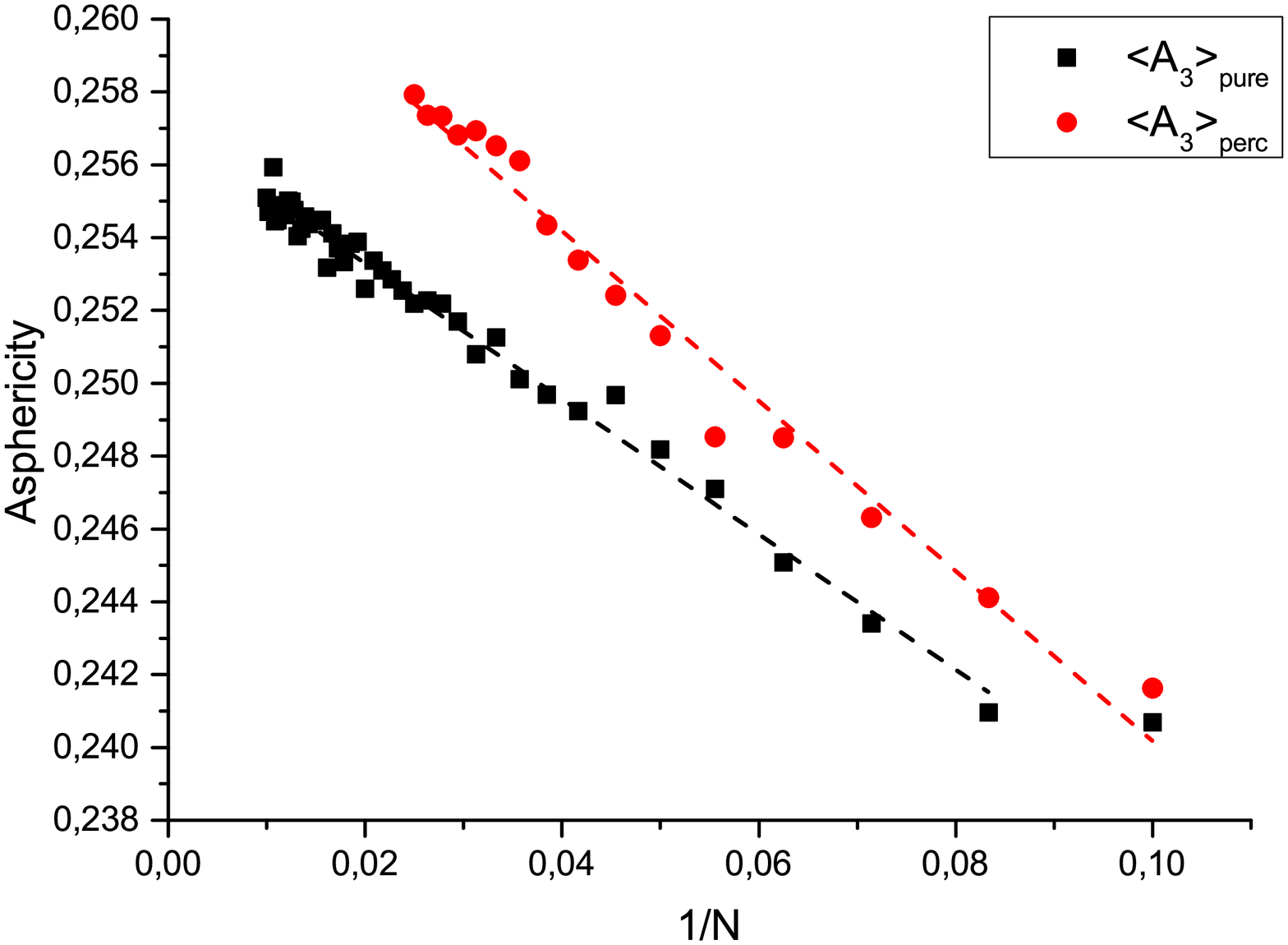}\includegraphics[width=73mm]{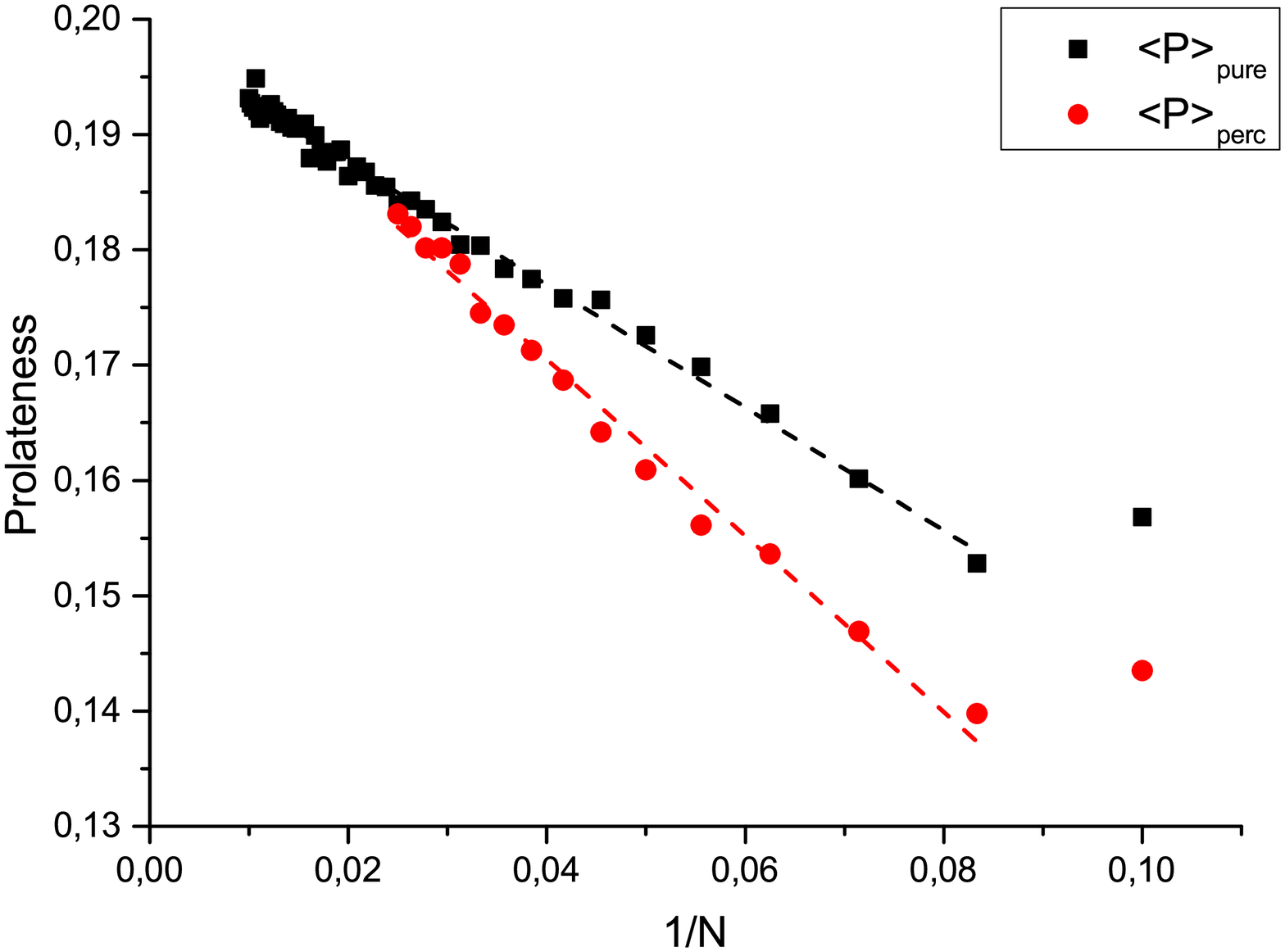}
\caption{ \label{fig:2} Aspericity on the left and  prolateness on the right as functions of $1/N$. Black squares: pure lattice, red circles:  percolation cluster.}
\end{center}
\end{figure}

Let us consider the  size ratios defined by Eqs. (\ref{g}) and (\ref{p}), which allow us to compare the effective extension of closed rings comparing to
 the linear chains. In Fig. \ref{fig:1} we present our simulation  data as functions of trajectory length $N$.
Note, that at small $N$ the size ratio of gyration radii of ring and linear trajectories on percolation cluster
$g_{{\rm perc}}(N)$ is smaller than that on the pure lattice $g_{{\rm pure}}(N)$, whereas it tends to increase faster with $N$.
On the other hand,  the ratio of the spanning radii on percolation cluster $p_{{\rm perc}}(N)$ is always larger than that on the pure lattice
$p_{{\rm pure}}(N)$.

For finite trajectory lengths $N$ considered in our study, the values of size ratios considerably differ from those for infinitely long structures.
The size ratio estimates can be obtained by using a least-square fitting of the form:
\begin{eqnarray}
&&g_{{\rm pure}}(N)=g_{{\rm pure}}+A_1/N, \,\, g_{{\rm perc}}(N)=g_{{\rm perc}}+A_2/N,\\
&&p_{{\rm pure}}(N)=p_{{\rm pure}}+B_1/N, \,\, p_{{\rm perc}}(N)=p_{{\rm perc}}+B_2/N,
\end{eqnarray}
with $A_1$, $A_2$, $B_1$, $B_2$ being constants.
 Results of the approximation are given in table \ref{t1}. Note that in pure environment our estimates are in good agreement with previous results
 even though we consider very short trajectories.

Next, we turn our attention to evaluating  the set of shape parameters.  In Fig. \ref{fig:2}
we present our simulation  data as functions of length $N$.
 The  estimates can again be obtained by using a least-square fitting of the form:
\begin{eqnarray}
&&{\langle A_3(N) \rangle}_{{\rm pure}}={\langle A_3 \rangle}_{{\rm pure}}+C_1/N, \,\,
\overline{\langle A_3(N) \rangle}_{{\rm perc}}=\overline{\langle A_3 \rangle}_{{\rm perc}}+C_2/N,\\
&&{\langle P(N) \rangle}_{{\rm pure}}={\langle P\rangle}_{{\rm pure}}+D_1/N,
\,\, \overline{\langle P(N) \rangle}_{{\rm perc}}=\overline{\langle P \rangle}_{{\rm perc}}+D_2/N,
\end{eqnarray}
with $C_1$, $C_2$, $D_1$, $D_2$ being constants.
 Results of the approximation are given in table \ref{t1}.
Again our results on pure lattice are in good agreement with previous calculations. The value of prolateness 
$\overline{\langle P \rangle}_{{\rm perc}}$ is almost not modified by presence of disorder, whereas asphericity $\overline{\langle A_3 \rangle}_{{\rm perc}}$ is larger as comparing with the case of pure lattice and thus quantitatively describes the increase of asymmetry of typical closed ring trajectory. 

\section{Conclusions}\label{conc}

In the present work we analyzed the conformational properties of closed flexible polymers in disordered environment within the
lattice model of SAWs on percolation clusters.  The complicated  structure of percolation cluster
captures  in particular  the irregular  gel structure with  the multiple dangling ends \cite{Whytock},
 and thus the proposed model may describe  the situation
when the ring polymers are placed in cross-linked polymer gels.

 The size and shape properties of a specified polymer conformation
 are characterized in terms of the gyration tensor $\bf Q$.   The rotationally invariant quantities constructed as combinations of
components of $\bf Q$, such as gyration radius ${\langle R_{g}^2\rangle}$,
averaged asphericity $\langle A_3 \rangle$ and prolateness $\langle P \rangle$
are directly obtained in numerical simulations with application of pivot algorithm.
All the shape characteristics increase gradually with increasing the length of SAW trajectory. 
The presence of disorder makes the longer conformations to be more elongated and asymmetric. 
 Our results quantitatively indicate the change of the size and shape parameters of typical closed ring 
  conformations relative to the obstacle-free case.

\end{document}